\documentclass[conference, letterpaper]{IEEEtran}
\usepackage{epsfig, amsmath, cite}
\usepackage{algorithmic, algorithm}
\usepackage{caption,subcaption}

\begin{document}

\title{Forward Delay-based Packet Scheduling Algorithm for Multipath TCP}

\author{
  Tuan-Anh Le\\
  Thu Dau Mot University\\
  Binh Duong, Vietnam\\
  Email: letuanh@tdmu.edu.vn
  \and
  Loc X. Bui\\
  Tan Tao University and Misfit Wearables Corp.\\
  Ho Chi Minh City, Vietnam\\
  Email: locbui@ieee.org
}

\maketitle

\begin{abstract}
Multipath TCP (MPTCP) is a transport layer protocol that allows network devices to transfer data over multiple concurrent paths, and hence, utilizes the network resources more effectively than does the traditional single-path TCP. However, as a reliable protocol, MPTCP still needs to deliver data packets (to the upper application) at the receiver in the same order they are transmitted at the sender. The out-of-order packet problem becomes more severe for MPTCP due to the heterogeneous nature of delay and bandwidth of each path. In this paper, we propose the forward-delay-based packet scheduling (FDPS) algorithm for MPTCP to address that problem. The main idea is that the sender dispatches packets via concurrent paths according to their estimated forward delay and throughput differences. Via simulations with various network conditions, the results show that our algorithm significantly maintains in-order arrival packets at the receiver compared with several previous algorithms.
\end{abstract}

\section{Introduction}\label{sec:introduction}

It is common nowadays that the network devices are equipped with multiple access interfaces, and hence, will potentially have multiple concurrent paths through the Internet's infrastructure. For instances, a personal computer may have both Ethernet and Wi-Fi interfaces. Similarly, a smart phone usually has multiple radio interfaces, such as cellular (3G/4G), Wi-Fi, Bluetooth, etc. On the other hand, the traditional TCP protocol only assumes one data flow through the network, and hence, may not utilize the network resources efficiently. To overcome that, the Internet Engineering Task Force (IETF) has standardized the Multipath TCP (MPTCP) protocol (see \cite{rfc6182, rfc6356, rfc6824}) as a major extension of the single-path TCP protocol. It allows simultaneous data transfer between end-to-end hosts across different paths, and hence, has the capability to increase the end-to-end goodput.

In order to utilize multiple paths in MPTCP, a packet scheduler is required; its responsibility is to decide which packet to be dispatched via which path. On the other hand, as a reliable protocol, MPTCP still needs to deliver data packets (to the upper application) at the receiver in the same order they are transmitted at the sender. Due to the heterogeneous nature of delay and bandwidth of each path, data packets may arrive out-of-order at the receiver and have to be re-ordered. Thus, a well designed packet scheduling algorithm will lead to a less re-ordering effort (both buffer-wise and computation-wise) at the receiver, and prevent performance degradation of MPTCP.

In this work, we propose the forward-delay-based packet scheduling (FDPS) algorithm for MPTCP to address the problem of out-of-order received packets described above. In particular, our main contributions can be summarized as follows:
\begin{itemize}
\item We present a technique to calculate the differences in forward delays between paths that \emph{does not require clock synchronization} between source host and destination host. This technique is of interest on its own, and may be useful to other future developments related to MPTCP.
\item Based on the above technique, we propose our FDPS algorithm to address the problem of out-of-order received packets. The main idea is to schedule packets according to the estimated delay and bandwidth differences between paths in the forward direction.
\item We implement and evaluate our proposed algorithm in ns-2 \cite{ns-2} and via the Reorder Buffer-occupancy Density (RBD) metric, respectively. Our extensive simulation results (under various network conditions) show that our algorithm achieve the more fraction of in-order received packets than do the previous algorithms.
\end{itemize}

The remainder of the paper is organized as follows. We review the previous works related to MPTCP and packet scheduling in Section \ref{sec:related_work}. In Section \ref{sec:scheduling_alg}, we describe our proposed FDPS algorithm for MPTCP, and present performance evaluation results in Section \ref{sec:perf_evaluation}. We then conclude our work in Section \ref{sec:conclusion}.

\section{Related Work}\label{sec:related_work}

Various congestion control algorithms have been proposed for MPTCP, e.g., \cite{kelvoi05,hanshaholsritow06, wisraigrehan11, rfc6356, penwallow13} to take advantage of multiple concurrent paths. In general, the MPTCP congestion control will send more data through the paths that have larger congestion window size and/or smaller round trip time (RTT). However, as mentioned in the Introduction section, the differences of bandwidth and delay between paths may cause out-of-order packets at the receiver. For instance, it has been shown \cite{singoetimschban12} that if two paths have a large difference in bandwidth and delay, MPTCP on these paths without a proper packet scheduling algorithm will be less effective than traditional single-path TCP.

Recently, several packet scheduling solutions have been proposed to address the out-of-order packet problem in MPTCP. For example, the multipath transmission control scheme (MTCS) \cite{tsachiparshi10} attempts to combine congestion control and scheduling to send in-order packets. The scheme uses a load sharing model to distinguish packet loss due to network congestion from the one due to wireless channel condition to find the optimal path. At the same time, it employs a feedback-control-based packet scheduling mechanism to maximize the number of data packet sent to the receiver in a timely manner without losing their order.
Kim et al.'s algorithm \cite{kimohlee12} (called Kim-2012 onward in this paper) performs in-order packet scheduling by dispatching packets on each path based on RTT after successfully sending a data packet and receiving acknowledgment (ACK) for that packet. However, the algorithm uses RTT/2 as an estimate for the forward delay, and as illustrated in Fig. \ref{fig:forward_backward_delay} it may not be a good estimation due to the asymmetric delay nature of forward and backward paths.

Although our focus is to estimate the forward delay difference between concurrent paths and not the forward delay itself, we still would like to briefly survey the literature on one-way delay estimation without clock synchronization. Tsuru et al. \cite{tsutakoie02} have proposed a method to estimate clock offset (between two hosts) for the case in which the forward and backward paths have different bandwidths, but they still assume symmetric propagation and transmission delays on both paths. On the other hand, the method proposed by Choi and Yoo \cite{choyoo05} does not require any symmetric or synchronization assumption, but it requires the RTT measurements at both sender and receiver, and its accuracy strongly depends on the delays of first packets in both directions.

\section{Forward Delay-based Packet Scheduling (FDPS) Algorithm}
\label{sec:scheduling_alg}

In this section, we present the details of our proposed forward-delay-based packet scheduling (FDPS) algorithm. The algorithm consists of two parts: i) estimating the forward delay differences between paths, and ii) selecting data to send via a path when the congestion window is available on that path.

\subsection{Estimating forward delay differences}

\begin{figure}[t]
    \begin{center}
    \includegraphics[width=6.5cm]{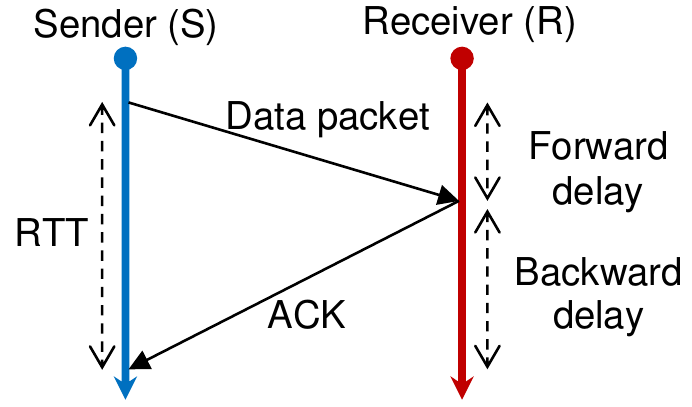}
    \end{center}
    \caption{Forward and backward delays}
    \label{fig:forward_backward_delay}
\end{figure}

Forward delay is defined as the delay in forward direction from sender to receiver. Similarly, backward delay is the delay in backward direction (from receiver back to sender), and the total of forward and backward delays is the round-trip time (RTT) (see Fig. \ref{fig:forward_backward_delay} for an illustration).
The problem of estimating forward delay on each path is not trivial in practice because the clocks at sender and receiver are often not synchronized. Nevertheless, in this section, we show that it is possible to estimate the forward delay differences between paths even without clock synchronization of sender and receiver.

\begin{figure}[t]
    \begin{center}
    \includegraphics[width=6.5cm]{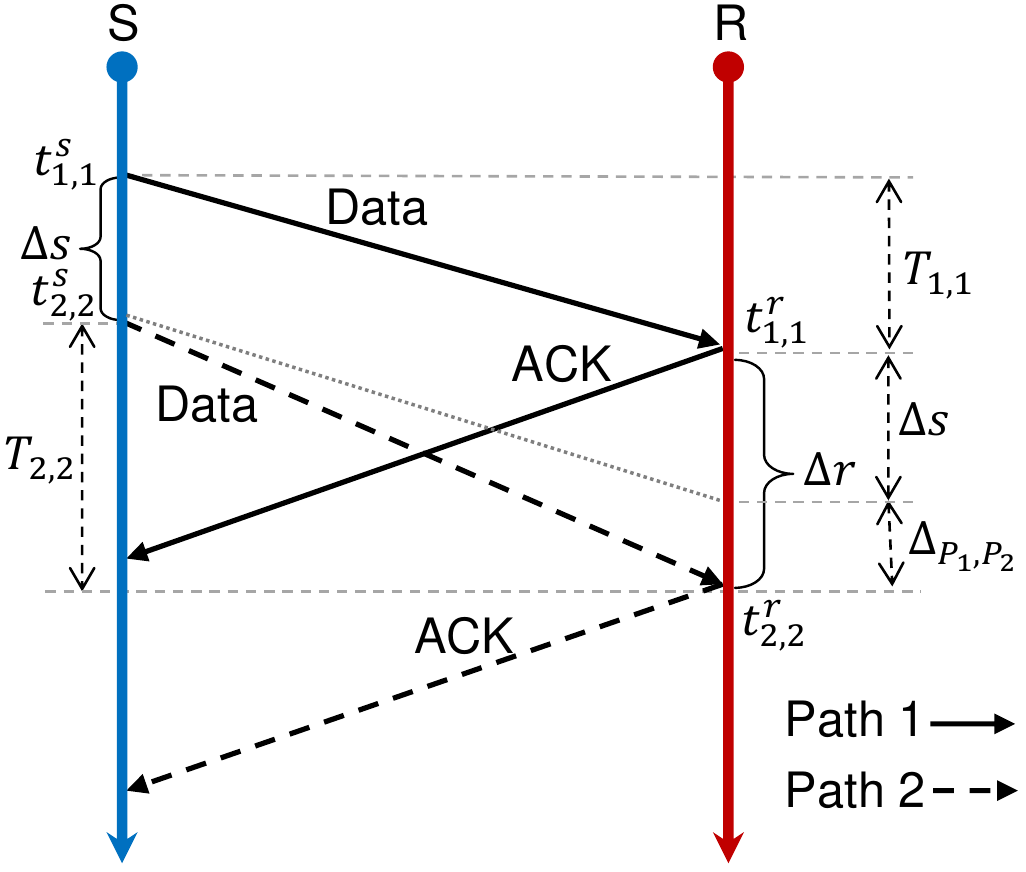}
    \end{center}
    \caption{Calculating the forward delay difference}
    \label{fig:forward_delay_diff}
\end{figure}

Our technique is illustrated in Fig. \ref{fig:forward_delay_diff},
in which the sender (S) sends two packets via two different paths to the receiver (R). In particular, at time $t^s_{1,1}$, S sends packet 1 via Path 1 to R, and that packet reaches R at time $t^r_{1,1}$. At time $t^s_{2,2}$, S sends packet 2 via Path 2 to R, and that packet reaches R at time $t^r_{2,2}$. We define the sending time difference $\Delta s$ between two packets as:
\begin{equation}
\Delta s = t^s_{2,2} - t^s_{1,1}, \quad \Delta s \in R, \label{eqn:delta_s}
\end{equation}
and similarly, the receiving time difference $\Delta r$ between two packets:
\begin{equation}
\Delta r = t^r_{2,2} - t^r_{1,1}, \quad \Delta r \in R. \label{eqn:delta_r}
\end{equation}
Let $\Delta T$ be the difference in clock time between S and R, i.e., $\Delta T = clk_S - clk_T,$ $\Delta T \in R.$ ($\Delta T$ is possibly unknown to both S and R.) Then the forward delay of packet 1 on Path 1 is:
\begin{equation}
T_{1,1} = t^r_{1,1} - t^s_{1,1} + \Delta T. \label{eqn:forward_delay_1}
\end{equation}
Similarly, the forward delay of packet 2 on Path 2 is:
\begin{equation}
T_{2,2} = t^r_{2,2} - t^s_{2,2} + \Delta T. \label{eqn:forward_delay_2}
\end{equation}
Therefore, the forward delay difference between Path 1 and Path 2 is estimated as:
\begin{align}
\Delta_{P_1,P_2} &= T_{1,1} - T_{2,2} \nonumber \\ 
      &= t^r_{1,1} - t^s_{1,1} + \Delta T - t^r_{2,2} + t^s_{2,2} - \Delta T \nonumber \\
      &= t^s_{2,2} - t^s_{1,1} - t^r_{2,2} + t^r_{1,1} \nonumber \\
      &= \Delta s - \Delta r. \label{eqn:forward_delay_diff}
\end{align}
Note that the sender S is able to carry out the above estimation after receiving ACKs of both packets from both paths. The sending timestamps $t^s_{i,j}$ can be put into the data packet option and then piggybacked to the sender via the ACK packet option, while the receiving timestamps $t^r_{i,j}$ can be put into the ACK packet option.

\subsection{Scheduling data packets}

In order to schedule data packets, we first need to find the path with shortest forward delay. Particularly, let $\mathcal{P} = \{P_1, P_2, \cdots, P_N\}$ be the set of MPTCP concurrent paths, then the path with shorted forward delay $P_{i^*}$ can be determined by {\bf Algorithm \ref{alg:SFDP}}.
\begin{algorithm}[ht]
  \caption{Finding shortest forward-delay path}
  \label{alg:SFDP}
\begin{algorithmic}
  \STATE {\bfseries Input:} a set of paths $\mathcal{P} = \{P_1, P_2, \cdots, P_N\}$
  \STATE {\bfseries Output:} $P_{i^*}$ as the path with shortest forward-delay
  \STATE
  \FOR{$i = 1$ to $N$}
    \STATE NegCount$[i]$ $\leftarrow$ $0$
  \ENDFOR
  \FORALL{$P_i, P_j \in \mathcal{P}$, $i \neq j$}
    \STATE Calculate $\Delta_{P_i,P_j}$ using Equation (\ref{eqn:forward_delay_diff})
    \IF {$\Delta_{P_i,P_j} < 0$}
      \STATE NegCount$[i]$ $\leftarrow$ NegCount$[i]$ $+$ $1$
    \ELSIF {$\Delta_{P_i,P_j} > 0$}
      \STATE NegCount$[j]$ $\leftarrow$ NegCount$[j]$ $+$ $1$
    \ELSE
      \STATE Generate a Bernoulli sample $s \in \{0,1\}$
      \IF {$s = 0$}
        \STATE NegCount$[i]$ $\leftarrow$ NegCount$[i]$ $+$ $1$
      \ELSE
        \STATE NegCount$[j]$ $\leftarrow$ NegCount$[j]$ $+$ $1$
      \ENDIF
    \ENDIF
  \ENDFOR
  \STATE {\em /* $P_{i^*}$ is the path with maximum NegCount */}
  \STATE $i^* := \arg\max \mbox{NegCount}[i]$
\end{algorithmic}
\end{algorithm}

The main idea of {\bf Algorithm \ref{alg:SFDP}} is fairly simple: the forward delay difference $\Delta_{P_{i^*},P_j}$ between $P_{i^*}$ and any other $P_j \in \mathcal{P}$ has to be less than or equal to zero. Thus, the algorithm does the comparison for all pairs of paths, and for each path it keeps track of the number of times when the difference is negative for that path (stored in the NegCount variable, with random tie-breaking). Then $P_{i^*}$ is the path with largest value of NegCount.

\begin{figure}[tb]
    \begin{center}
    \includegraphics[width=7cm]{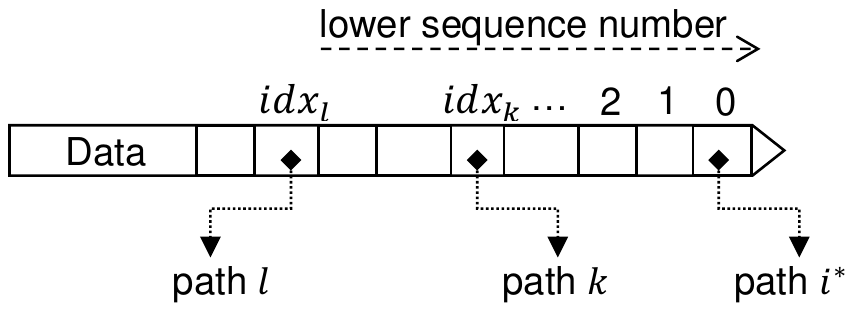}
    \end{center}
    \caption{Dispatching packets based on forward delay differences}
    \label{fig:packet_dispatching}
\end{figure}

After the path with shortest forward delay is determined, the data packets will be scheduled. Suppose that the MPTCP sender uses a sending buffer shared among sub-flows, and data coming from the application layer is enough to fill in the buffer. Let the data in the sending buffer be partitioned into packets, each packet has the size equal to the MSS (message segment size) of TCP paths. Then the scheduler chooses data packets with lower sequence numbers to send on paths with smaller forward delays. In particular, when the sub-flow on path $i$ requires data to transfer, the scheduler will choose data packet at index $idx_i$ in the sending buffer according to the following formula:
\begin{equation} \label{eq:index_i}
idx_i = \Delta_{P_i,P_{i^*}} \times \frac{x_{i^*}}{\mbox{MSS}},
\end{equation}
where $x_i$ is the average throughput (byte/s) path $i$.
Note that path $i^*$ always picks the data packet at index $0$. After one data packet is sent, the other packets with higher indices will be shifted to fill up the buffer. An illustration of the algorithm is presented in Fig. \ref{fig:packet_dispatching}.

We would like to emphasize that our packet scheduling algorithm is a Pull mechanism (as opposed to Push or Hybrid push-pull mechanisms \cite{singoetimschban12}). That is, generated data packets are stored in the sending buffer and allocated to the sub-flows only when the sub-flows have an open window to transmit data (e.g., after receiving an ACK).

We conclude this section with some discussions about the intuition behind our proposed algorithm. One can see that $x_{i^*} / \mbox{MSS}$ is the normalized rate (in packet per second) of path $i^*$, and hence, the product $\Delta_{P_i,P_{i^*}} \times x_{i^*} / \mbox{MSS}$ may be interpreted as the ``bandwidth-delay product difference'' between paths $i$ and $i^*$. Therefore, there is roughly enough room on path $i^*$ to send $\Delta_{P_i,P_{i^*}} \times x_{i^*} / \mbox{MSS}$ packets before getting to path $i$, and thus, the packet to send on path $i$ should be right after those packets, i.e., at the index presented in Equation (\ref{eq:index_i}).

\section{Performance Evaluations}\label{sec:perf_evaluation}

In this section, we investigate the performance of our FDPS algorithm and compare it with the performances of MTCS \cite{tsachiparshi10}, Kim-2012 \cite{kimohlee12}, and FIFO (first-in-first-out, default in MPTCP) in terms of Reorder Buffer-occupancy Density (RBD) and Reorder Density (RD) metrics \cite{rfc5236}, both are measured at the receiver. By definition, RBD is the normalized histogram of the occupancy of a hypothetical buffer (at the receiver) needed to re-sort out-of-order packets, while RD is defined as the distribution of displacements of packets from their original positions, normalized with respect to the number of packets. Note that RBD[$0$] (the value of RBD at index $0$) represents the density of in-order packets that arrive at the receiver and can be delivered to upper application immediately without buffering. 

We would like to point out that although throughput (or goodput) is a popular performance metric to benchmark TCP-related algorithms, we do not consider goodput here. The reason is that, beside the reordering aspect, goodput also depends on many other factors such as data loss or retransmission mechanism. On the other hand, both RBD and RD capture the essential nature of the reordering problem, and hence, are chosen as our main performance metrics.

\begin{figure}[tb]
    \begin{center}
    \includegraphics[width=9cm]{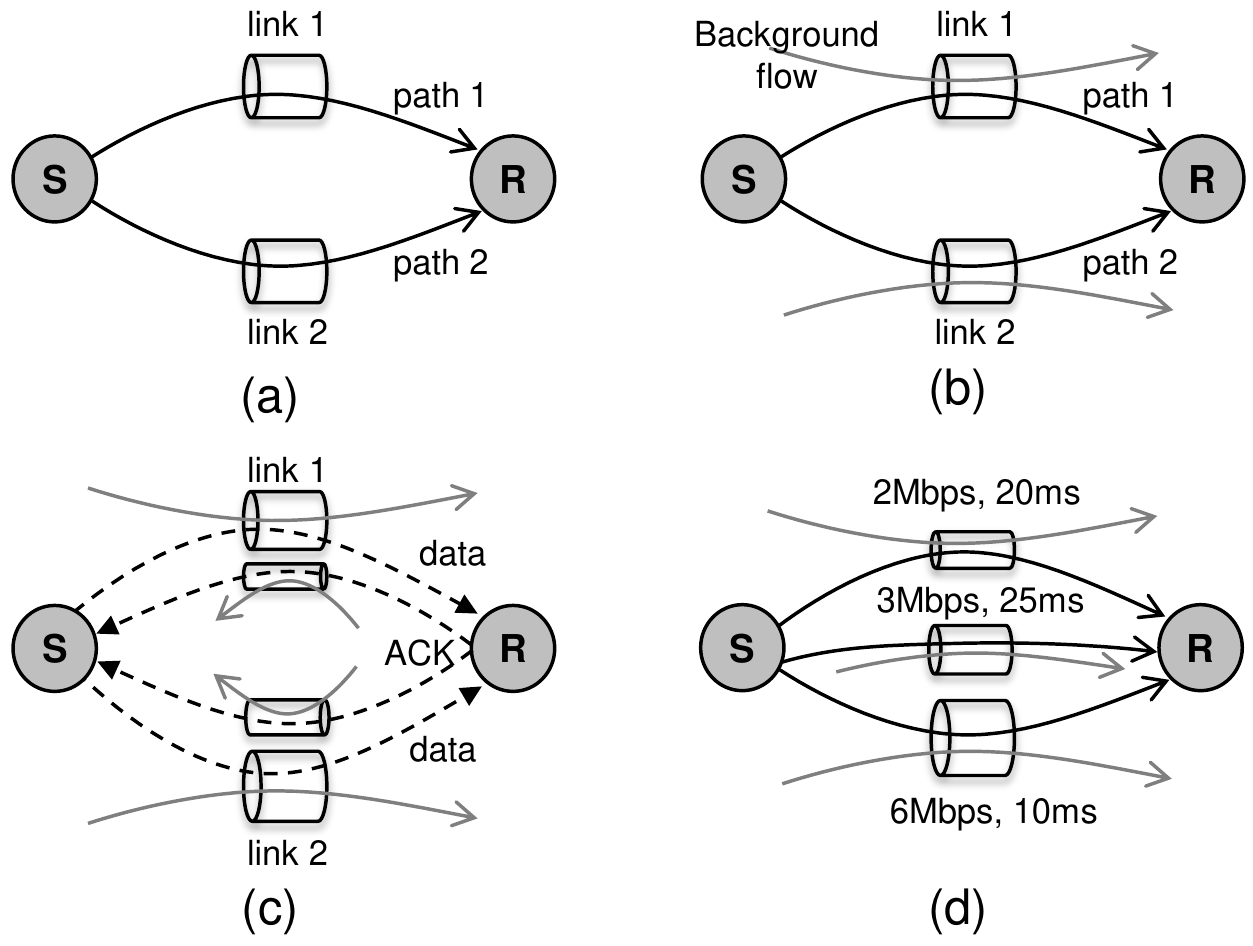}
    \end{center}
    \caption{Simulation topologies}
    \label{fig:topology}
\end{figure}

Our evaluations are based on topologies shown in Fig. \ref{fig:topology}
and simulated on NS-2.34 \cite{ns-2}. Simulation parameters are given in Table \ref{table:sim_para}.
\begin{table}[ht]
\begin{center}
\caption{Simulation parameters}
\label{table:sim_para}
\begin{tabular}{|p{4cm}|p{3.5cm}|}
\hline
\textbf{Parameter }& \textbf{Value} \\ \hline
Number of simulation runs for each\newline algorithm & 20 times \\ \hline
Simulation time for an experiment & 50 Seconds \\ \hline
Data packet size & 1000 Bytes \\ \hline
MSS & 934 Bytes \\ \hline
ACK option & Selective ACK \\ \hline
Queue management & RED \\ \hline
Router queue size & Bandwidth-Delay Product \\ \hline
Receive buffer & unlimit \\ \hline
Bandwidth side & 2 times of bottleneck link bandwidth \\ \hline
Delay side & 1 ms \\ \hline
Shortest forward-delay path update interval & every RTT \\ \hline
\end{tabular}
\end{center}
\end{table}
 The simulation experiments are carried out under various network conditions as presented in the following subsections.

\subsection{Two-path MPTCP without background traffic}
\begin{table}[ht]
\begin{center}
\caption{Experiment scenarios}
\label{table:setting_para}
\begin{tabular}{|p{1.5cm}|p{2.7cm}|p{2.7cm}|}
\hline
\textbf{Experiment\newline ID }& \textbf{Bottleneck link 1} & \textbf{Bottleneck link 2} \\ \hline
Exp. A1\newline (Fig. \ref{fig:topology}(a)) & BW = 4 Mbps,\newline  delay = 10 ms &  BW = 4 Mbps,\newline delay = 10 ms \\ \hline
Exp. A2\newline (Fig. \ref{fig:topology}(a)) & BW = 4 Mbps,\newline  delay = 10 ms &  BW = 4 Mbps,\newline delay = 30ms \\ \hline
Exp. A3\newline (Fig. \ref{fig:topology}(a)) & BW = 2 Mbps,\newline delay = 10 ms &  BW = 8 Mbps,\newline delay = 30ms \\ \hline
Exp. A4\newline (Fig. \ref{fig:topology}(b)) & BW = 4 Mbps,\newline delay = 10 ms &  BW = 4 Mbps,\newline delay = 30ms \\ \hline
Exp. A5\newline (Fig. \ref{fig:topology}(c)) & forward BW = 4 Mbps,\newline forward delay = 10 ms;\newline backward BW = 0.5 Mbps,\newline backward delay = 15 ms &  forward BW = 8 Mbps,\newline forward delay = 30 ms;\newline backward BW = 1 Mbps,\newline backward delay = 35 ms\\
\hline
\end{tabular}
\end{center}
\end{table}

\begin{figure}[tb]
    \begin{center}
    \includegraphics[width=9cm]{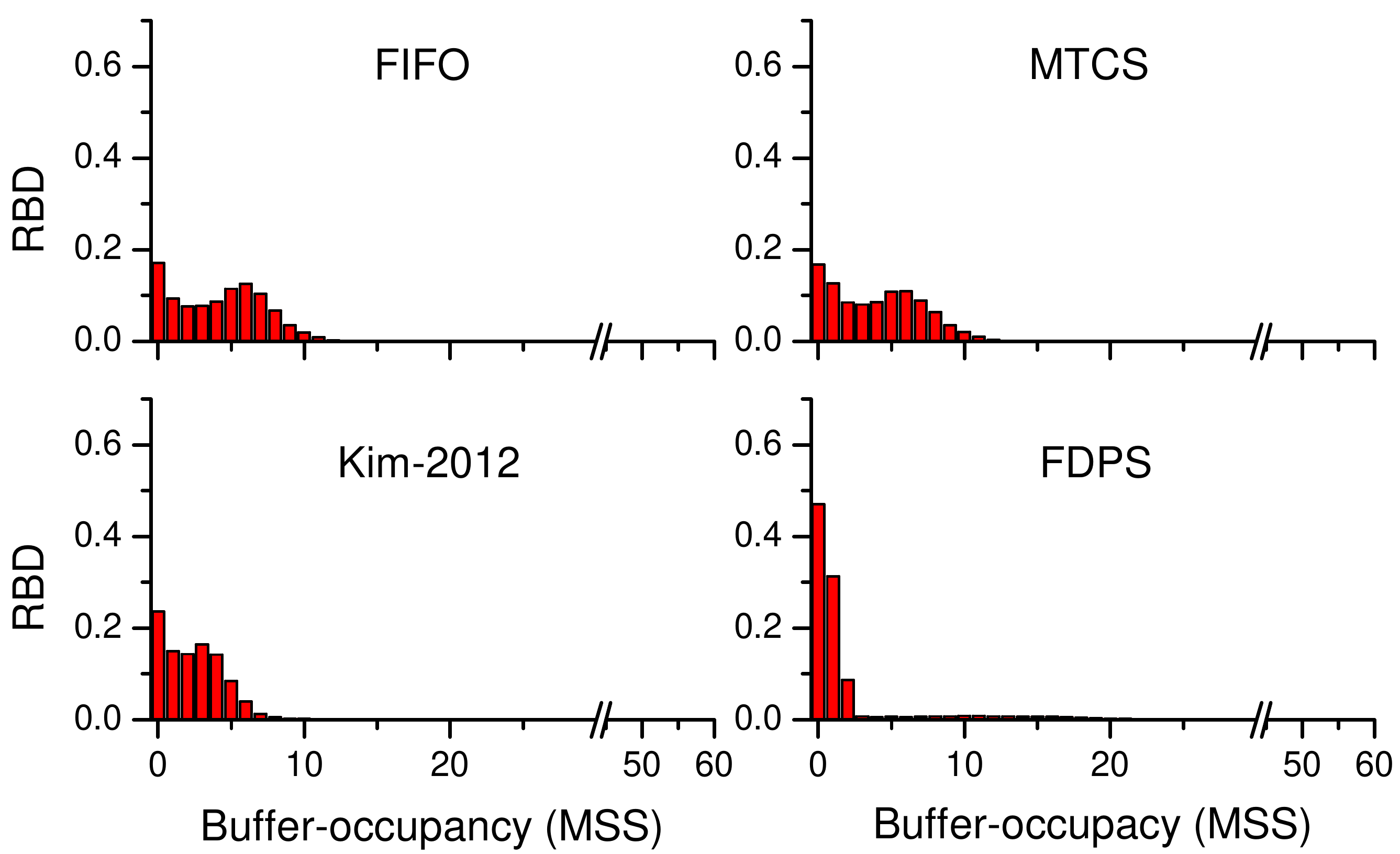}
    \end{center}
    \caption{RBD of the algorithms in Exp. A1}
    \label{fig:Exp_A1}
\end{figure}

In this section, we consider the topology shown in Fig. \ref{fig:topology}(a) in which a MPTCP flow runs on two separate paths without background traffic. The experiment (Exp.) A1 in Table \ref{table:setting_para} is used as the baseline case where two paths are identical (i.e., same propagation delay and bandwidth). Fig. \ref{fig:Exp_A1}
shows the RBDs of the considered packet scheduling algorithms in this case. One can see that even when two paths are identical, the out-of-order packet phenomenon still happens for all algorithms (the RBDs take values other than zero). Nevertheless, FDPS gets the best performance compared to all other algorithms; its RBD sharply concentrates around zero.

\begin{figure}[tb]
    \begin{center}
    \includegraphics[width=9cm]{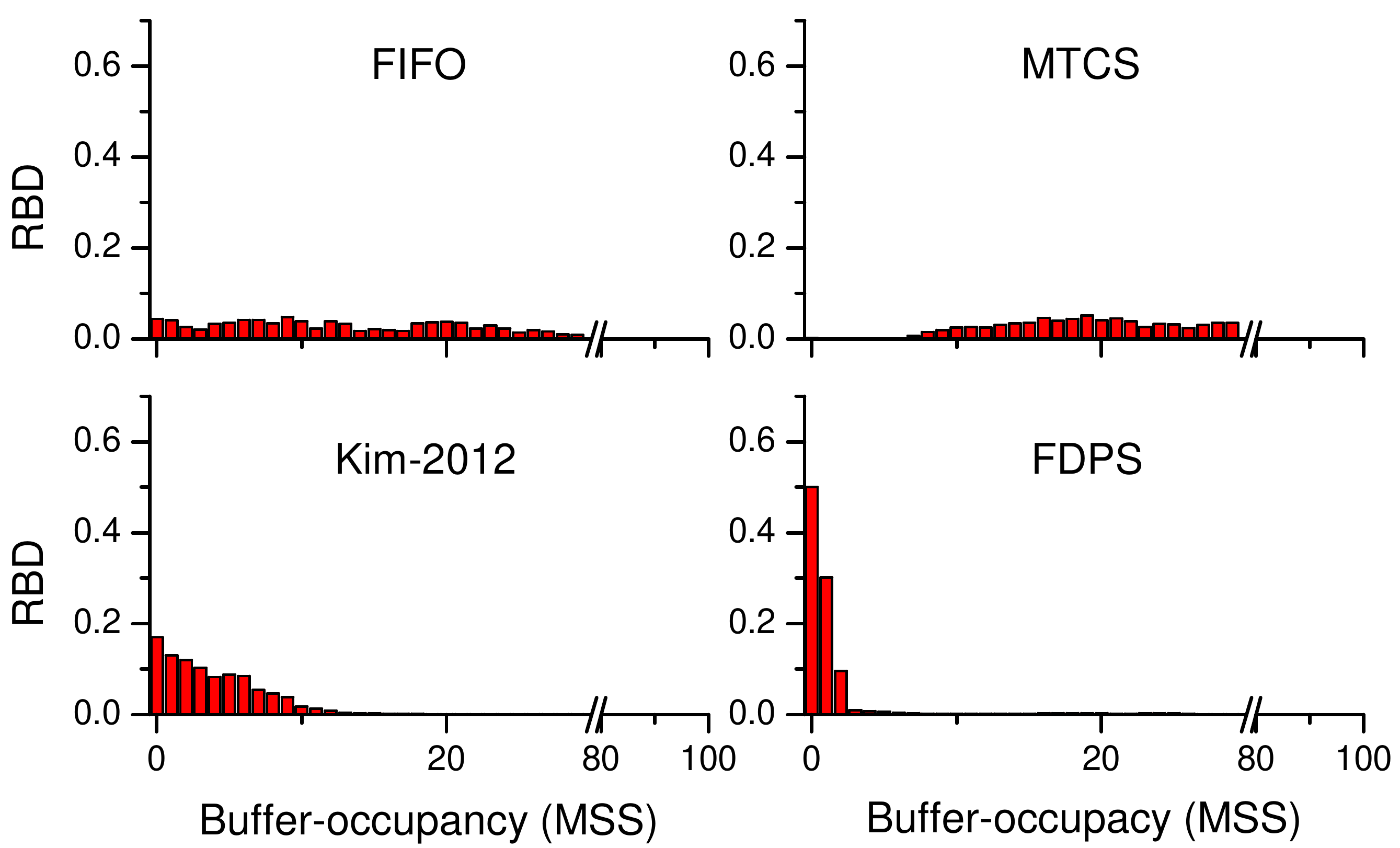}
    \end{center}
    \caption{RBD of the algorithms Exp. A2}
    \label{fig:Exp_A2}
\end{figure}

Next, in Exp. A2, the delay of path 2 is increased. The RBDs of the considered algorithms are shown in Fig. \ref{fig:Exp_A2}.
The performances of both FIFO and MTCS are degraded significantly. Kim-2012 performs much better than FIFO and MTCS in this case, but the performance of FDPS is still the best.

\begin{figure}[tb]
    \begin{center}
    \includegraphics[width=9cm]{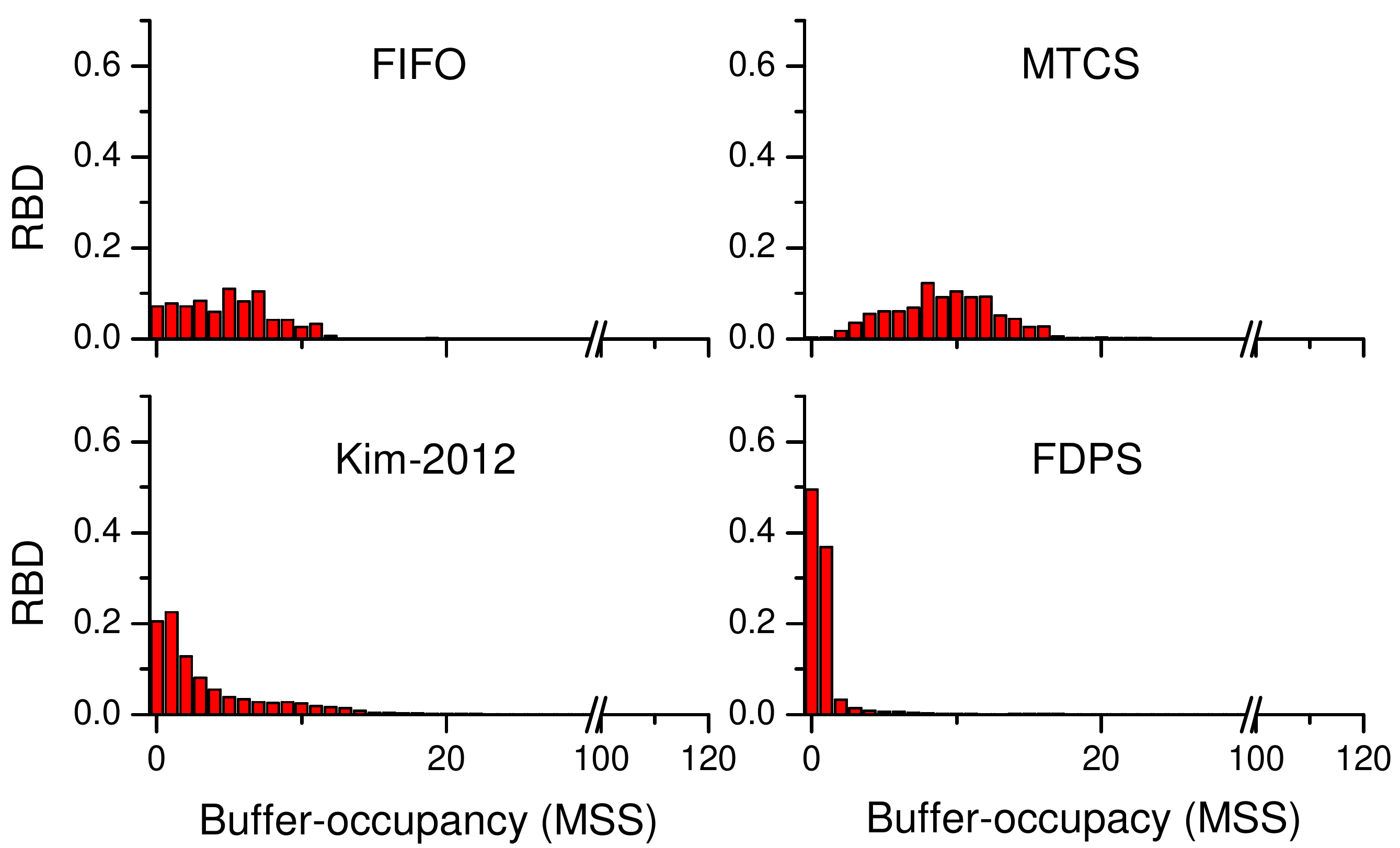}
    \subcaption{Reorder buffer density.}
    \includegraphics[width=9cm]{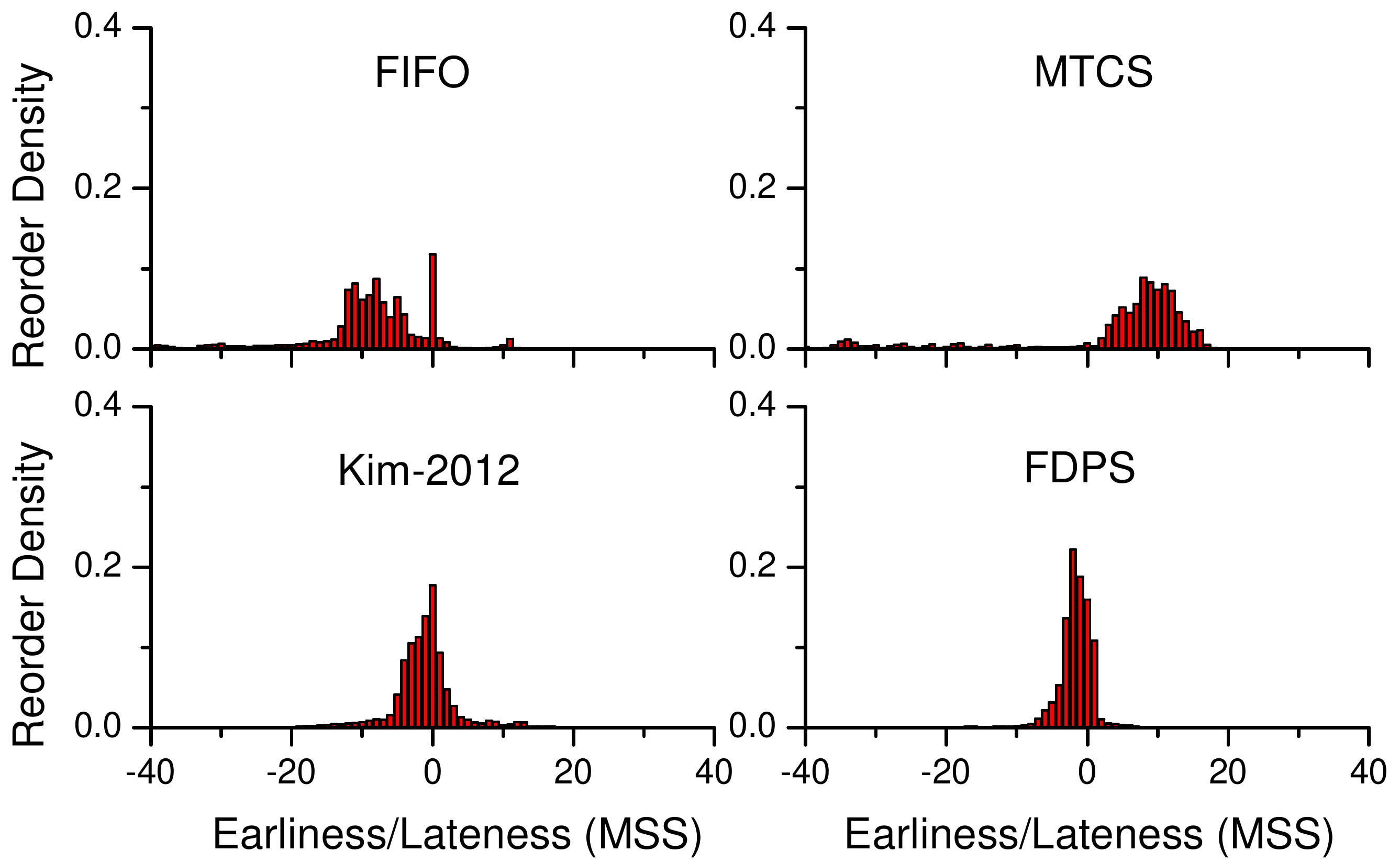}
    \subcaption{Reorder density.}
    \end{center}
    \caption{RBD (a) and RD (b) of the algorithms in Exp. A3}
    \label{fig:Exp_A3}
\end{figure}

Finally, in Exp. A3, we run simulation on two paths with different delay and bandwidth (path with lower bandwidth has smaller delay). Fig. \ref{fig:Exp_A3}(a) and Fig. \ref{fig:Exp_A3}(b)
show the RBDs and RDs of the considered algorithms, respectively. We still see that Kim-2012 is better than FIFO and MTCS, but FDPS achieves the best performance in both metrics. In particular, the packet reorder density of FDPS concentrates densely around zero, while those of the others spread widely.

\subsection{Two-path MPTCP with background traffic}

In practice, a MPTCP or TCP flow in Internet usually shares the bandwidth with one or more other flows in the forward and/or backward direction. To evaluate the efficiency of a packet scheduling algorithm when MPTCP sends data packets on such paths, we add one or more regular TCP flows to generate background traffic on each path.

\begin{figure}[tb]
    \begin{center}
    \includegraphics[width=9cm]{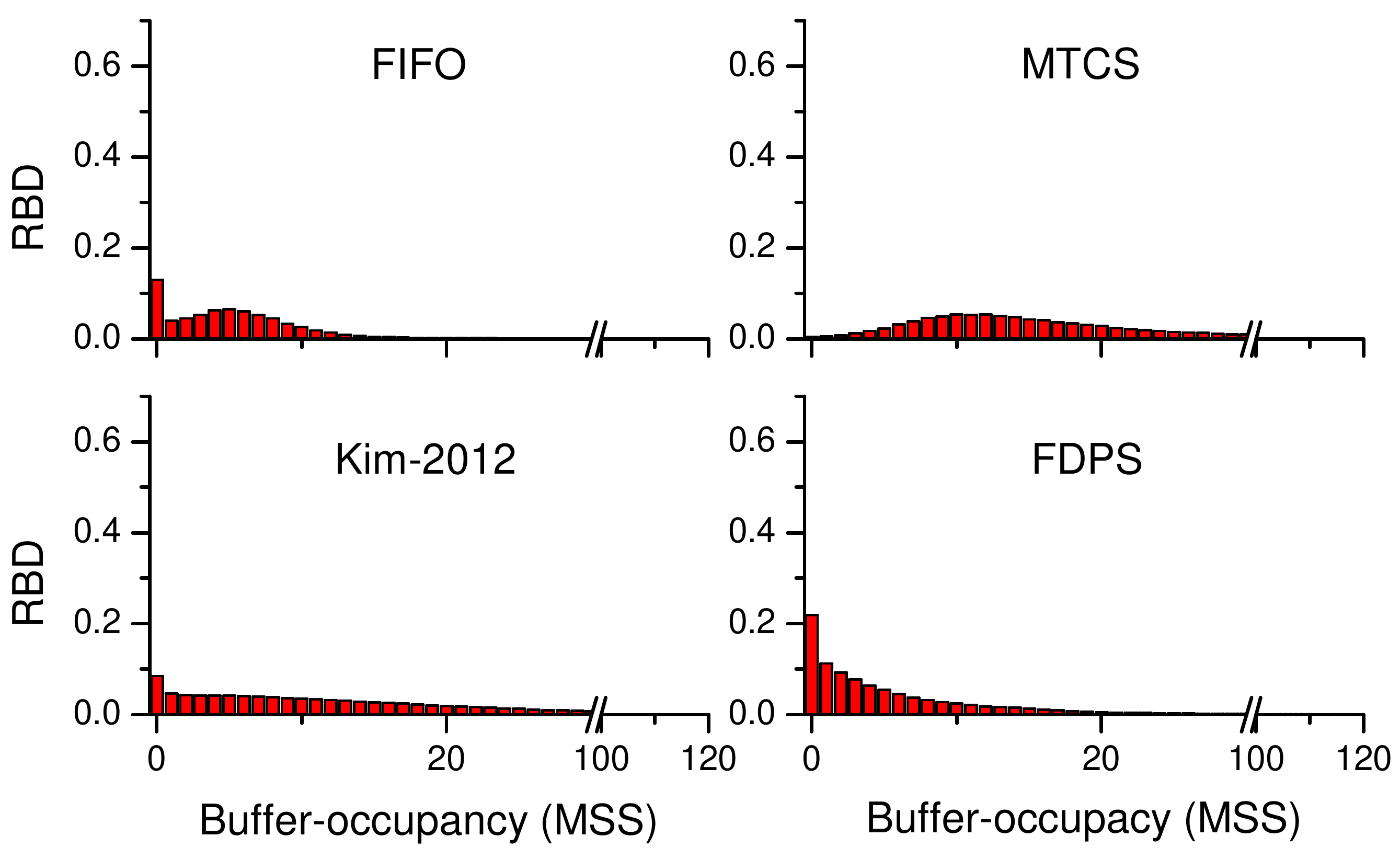}
    \end{center}
    \caption{RBD of the algorithms in Exp. A4}
    \label{fig:perf_with_bg_forward}
\end{figure}

In our fourth experiment, Exp. A4, the path configuration is the same as in Exp. A2, but there is background traffic in the forward direction as shown in Fig. \ref{fig:topology}(b). The results in Fig. \ref{fig:perf_with_bg_forward}
show that all algorithms suffer significant degradation in performance because the background flows make the RTT estimates less precise (due to large RTT variant) at the sender. Both Kim-2012 and MTCS yield poor performance than does the native FIFO algorithm because both are more sensitive to time variant than FIFO. However, FDPS still achieves the best performance in buffer occupancy.

\begin{figure}[tb]
    \begin{center}
    \includegraphics[width=9cm]{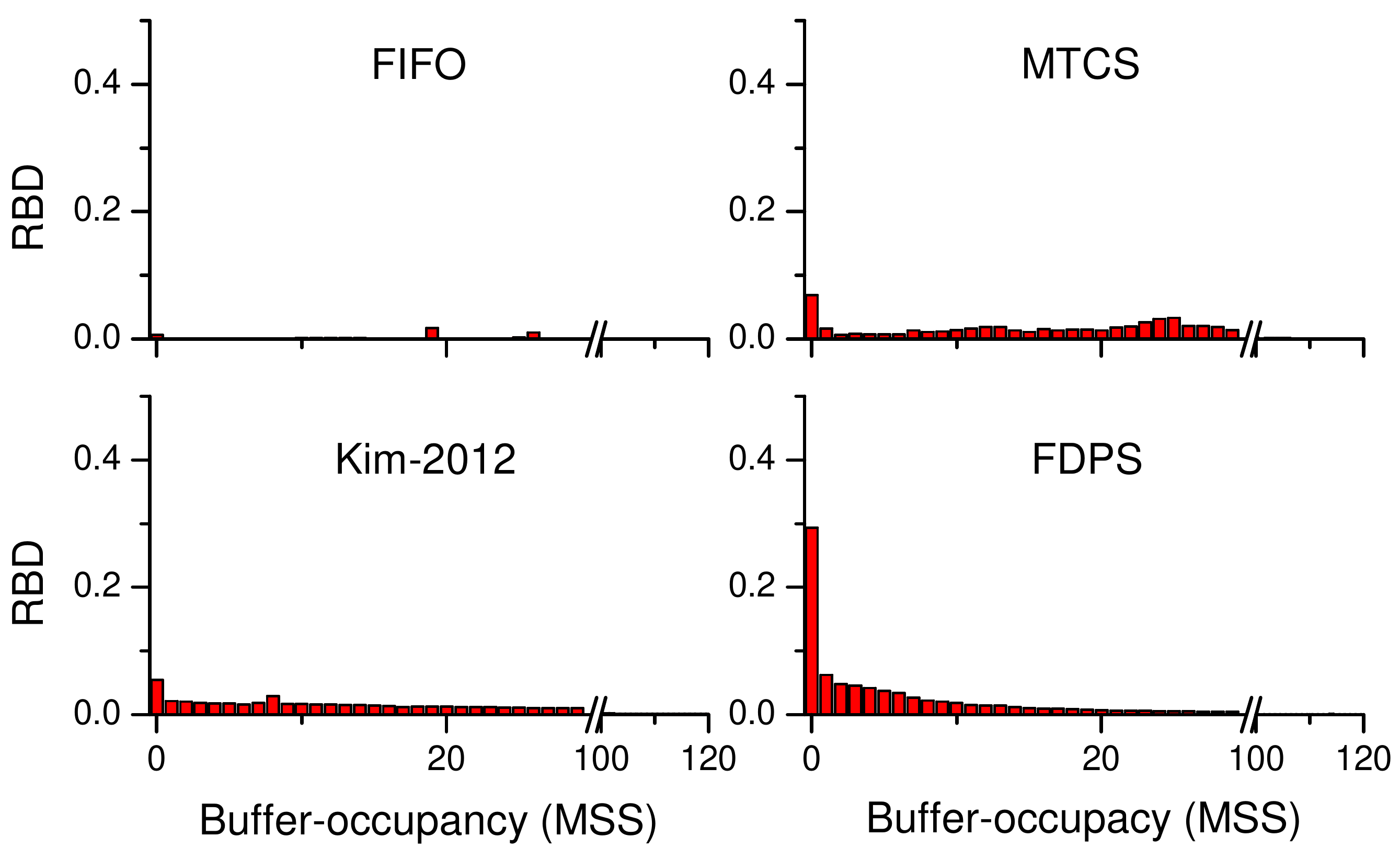}
    \end{center}
    \caption{RBD of the algorithms in Exp. A5}
    \label{fig:perf_with_bg_forward_backward}
\end{figure}

In the fifth experiment, Exp. A5, we use an asymmetric topology as shown in Fig. \ref{fig:topology}(c) with background flows in both forward and backward directions. Note that when regular TCP flows send data packets mixing with ACK packets (with small packet size), at the intermediate router(s) the ACK packets sometimes get to wait until data packets of background flows are forwarded first (this technique is called ACK compression \cite{rfc3449}). This might cause the measured RTT values become too small or too big, and hence, affect the performances of all considered algorithms, especially FIFO, as shown in Fig. \ref{fig:perf_with_bg_forward_backward}.
Nevertheless, FDPS still outperforms the others since it estimates only forward-delay differences between paths, thus not affected much by the backward paths.

\subsection{Three-path MPTCP}

\begin{figure}[ht]
    \begin{center}
    \includegraphics[width=9cm]{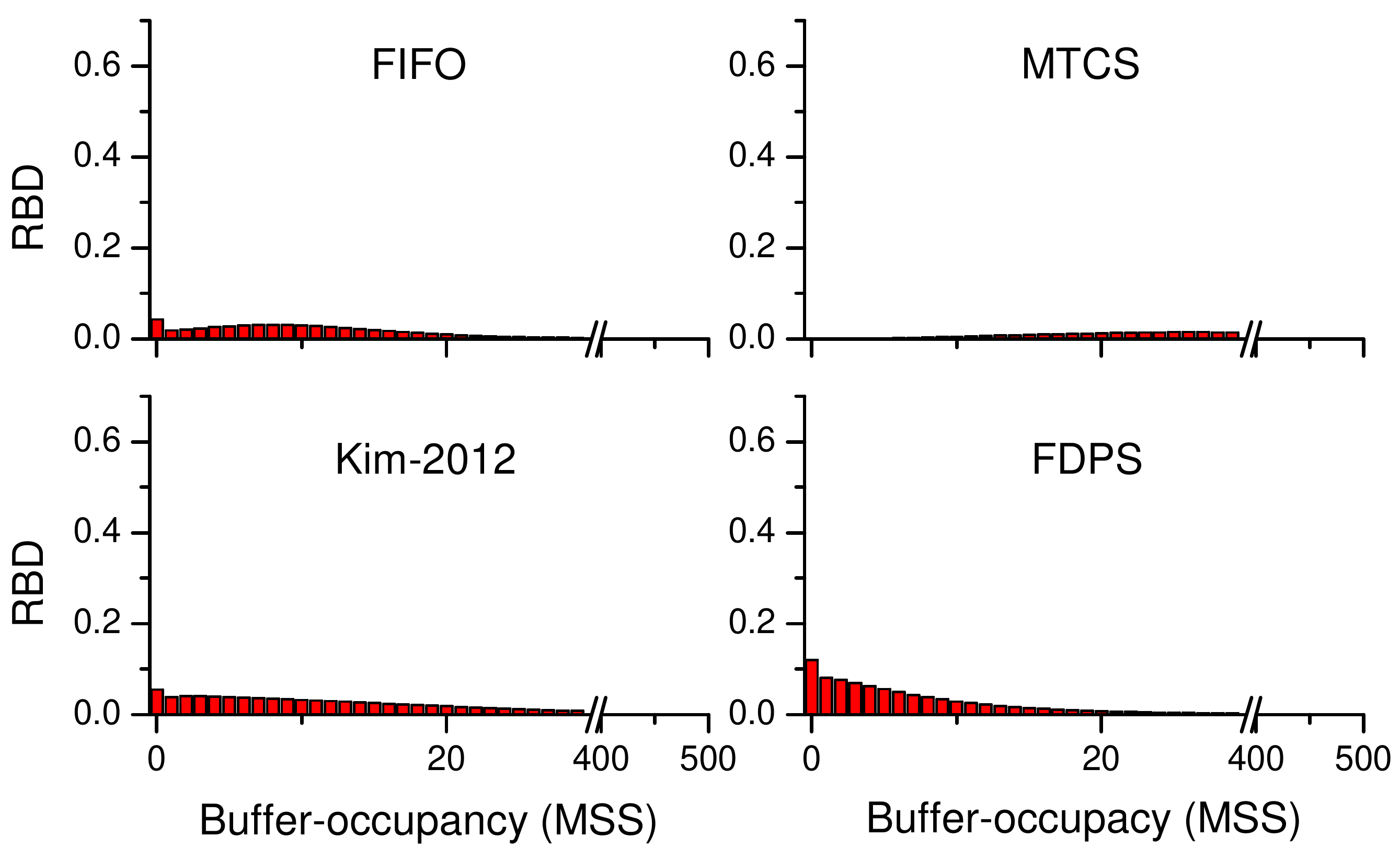}
    \end{center}
    \caption{RBD of the algorithms for the topology in Fig. \ref{fig:topology}(d)}
    \label{fig:three_paths}
\end{figure}

In the last experiment, we would like to evaluate the effectiveness of FDPS on a three-path MPTCP configuration. In particular, we run simulation on the topology in which paths have diverse values of bandwidth and delay (shown in Fig. \ref{fig:topology}(d)). Each path has a regular TCP flow generating background traffic. The results are shown in Fig. \ref{fig:three_paths}.
We can see that, compared to the two-path case, the performances of all considered algorithms get worse. It suggests that when the number of concurrent paths increases, the reordering problem gets more serious. In this case, FDPS again achieves the best reordering performance among the considered scheduling algorithms.

\subsection{Summary and discussions}

In summary, Table \ref{table:buffer_occupancy_mean} shows the means (and standard deviations) of buffer occupancy for all considered algorithms under the above experiments. Our proposed FDPS algorithm consistently gets the best buffer occupancy in all experiments, which suggests that FDPS is robust under various network conditions.
\begin{table}[ht]
\begin{center}
\caption{Mean of buffer occupancy (in MSS)}
\label{table:buffer_occupancy_mean}
\begin{tabular}{|l|l|l|l|l|}
\hline
\textbf{Experiment ID }& \textbf{FIFO} & \textbf{MTCS} & \textbf{Kim-2012} & \textbf{FDPS} \\ \hline
Exp. A1 & 4.5$\pm$0.16 &  4.3$\pm$0.18 & 2.8$\pm$0.08 & 2.2$\pm$0.02\\ \hline
Exp. A2 & 167.6$\pm$0.06 &  22.7$\pm$0.02 & 4.7$\pm$0.01 & 2.2$\pm$0.03\\ \hline
Exp. A3 & 105.5$\pm$0.04 &  10.0$\pm$0.01 & 5.4$\pm$0.01 & 2.6$\pm$0.02\\ \hline
Exp. A4 & 272.3$\pm$10.8 &  19.8$\pm$0.42 & 16.5$\pm$0.78 & 8.4$\pm$0.06\\ \hline
Exp. A5 & 1682$\pm$57.1 &  44.6$\pm$2.56 & 59.4$\pm$3.91 & 29.2$\pm$2.40\\ \hline
Exp. 3-paths & 790.1$\pm$67.2 &  68.7$\pm$1.96 & 32.4$\pm$2.02 & 24.5$\pm$1.14\\ \hline
\end{tabular}
\end{center}
\end{table}

Note that in our simulations, we use the unlimited reorder buffer size because we would like to capture all data packets at the receiver to have the correct RBD and RD measurements. In practice, that buffer size is usually finite, and hence, may cause packet loss if the out-of-order packet problem gets serious. Since FDPS achieves the best reorder buffer occupancy compared to previous solutions, the effect of finite reorder buffer size would be minimal if FDPS is used.


\section{Conclusions}\label{sec:conclusion}
In this paper, we introduce the forward-delay-based packet scheduling (FDPS) algorithm for Multipath-TCP to address the out-of-order received packet problem occurring when packets are transmitted via multiple concurrent paths. The algorithm is based on our novel technique of estimating the forward delay differences between paths. Our extensive simulation results show that our algorithm significantly reduces the reordering buffer occupation compared to the previous solutions to this problem.

\section*{Acknowledgment}
This research is funded by the Vietnamese National Foundation for Science and Technology Development (NAFOSTED) under grant number 102.02-2013.48.

\bibliographystyle{IEEEtran}

\bibliography{IEEEfull,mptcp}

\end{document}